\documentclass[preprint,prd,aps,tighten,nofootinbib,amssymb]{revtex4}
\usepackage{color}
\input{colordvi.tex}
\usepackage{amsmath}
\usepackage[dvips]{graphicx}
\usepackage{subfigure}
\usepackage{comment}
\newcommand{\beq}{\begin{equation}}
\newcommand{\eeq}{\end{equation}}
\newcommand{\beqa}{\begin{eqnarray}}
\newcommand{\eeqa}{\end{eqnarray}}

\newcommand{\bea}{\begin{eqnarray}}
\newcommand{\eea}{\end{eqnarray}}
\newcommand{\bear}{\begin{array}}
\newcommand {\eear}{\end{array}}
\newcommand{\bef}{\begin{figure}}
\newcommand {\eef}{\end{figure}}
\newcommand{\bec}{\begin{center}}
\newcommand {\eec}{\end{center}}

\newcommand{\la}{\left\langle}
\newcommand{\ra}{\right\rangle}

\def\GEV#1{10^{#1}{\rm\,GeV}}

\def\lrfp#1#2#3{ \left(\frac{#1}{#2} \right)^{#3}}

\begin{document}
\widetext
\draft

\begin{flushright}
KEK-TH-1625\\
UT-13-19\\
TU-934
\end{flushright}

\title{Moduli-Induced Axion Problem
}

\author{Tetsutaro Higaki}%
\affiliation{Theory Center, KEK, 1-1 Oho, Tsukuba, Ibaraki 305-0801, Japan}
\author{Kazunori Nakayama}
\affiliation{Department of Physics, University of Tokyo, Tokyo 113-0033, Japan}
\author{Fuminobu Takahashi}
\affiliation{Department of Physics, Tohoku University, Sendai 980-8578, Japan}

\date{\today}

\begin{abstract}
We point out that the cosmological moduli problem is not necessarily resolved
even if the modulus mass is heavier than $O(10)$\,TeV, contrary to the common wisdom. 
The point is that, in many scenarios where the lightest moduli fields are 
stabilized by supersymmetry breaking effects, those moduli fields  tend to mainly decay into almost
massless axions, whose abundance is tightly constrained by the recent Planck results.
We study the moduli-induced axion problem in concrete examples, and discuss 
possible solutions. The problem and its solutions are widely applicable to 
decays of  heavy scalar fields which dominate the energy density of the Universe, for instance,
the reheating of the inflaton.
\end{abstract}

\maketitle
\section{Introduction}
\label{sec:1}

The cosmological moduli problem~\cite{Coughlan:1983ci,de Carlos:1993jw} is one of the
most important issues in string theory and cosmology. The purpose of this paper
is to point out that the cosmological moduli problem is not necessarily resolved
even if the modulus mass is heavier than $O(10)$\,TeV, contrary to the common wisdom. 

In superstring theories~\cite{Polchinski:1998rq}, the so-called moduli fields appear at low energies through 
supersymmetric (SUSY) compactifications e.g., on a Calabi-Yau (CY) space~\cite{Candelas:1985en} 
as Kaluza-Klein zero-modes of 10-dimensional metric and $p$-form fields.
These moduli and their axionic superpartners are massless
at the perturbative level because of the shift symmetry; 
\begin{align}
T_{\rm moduli} \;\rightarrow\;  T_{\rm moduli}  + i \alpha,
\label{shiftsym}
\end{align}
which is regarded as a remnant of higher dimensional gauge symmetry,
with $\alpha$ being a real transformation parameter. 
In order to have a sensible low-energy theory, those moduli fields must be stabilized.
It is well known that many moduli fields are fixed simultaneously by the 
closed string flux backgrounds in extra dimensions, 
i.e. flux compactifications~\cite{Grana:2005jc, Blumenhagen:2006ci},
and most of the remaining moduli not fixed by the fluxes can be stabilized 
by instantons/gaugino condensations {\it a la} KKLT~\cite{Kachru:2003aw}.

During inflation, some of those moduli fields, especially relatively light ones, are likely
deviated from the low-energy minima. They will start coherent oscillations with a large amplitude 
at some time after inflation, and soon dominate the energy density of the Universe. 
If its mass is of order the weak scale, it typically
decays during the big bang nucleosynthesis (BBN), thus spoiling the overall agreement between 
BBN theory and light element observations.
If the modulus mass is much lighter, the situation  becomes worse;
the modulus abundance can easily exceed the dark matter
abundance or its decay may produce too much X-rays or gamma-rays~\cite{Asaka:1999xd}.
This is the notorious cosmological moduli problem~\cite{Coughlan:1983ci}.
Among many solutions proposed so far, the simplest one is to assume a heavy modulus mass;
the BBN bound is relaxed significantly if the modulus mass is heavier than several tens TeV.\footnote{
See Refs.~\cite{other-sol} for other solutions.}

The heavy moduli scenario does not necessarily lead to a successful cosmology.
It was pointed out in Refs.~\cite{moduli,Dine:2006ii,Endo:2006tf} that gravitinos are generically
produced by the modulus decay if kinematically allowed, and that gravitinos thus produced affect the light element abundance
and/or produce too many lightest supersymmetric particles (LSPs), even if the modulus decays before BBN.
This is known as the moduli-induced gravitino problem. There are several ways to avoid the problem. 
For instance, no gravitinos are produced if the decay is kinematically forbidden. This requires a heavy gravitino mass 
comparable to or heavier than the modulus. Alternatively, 
even if many gravitinos are produced by the modulus decay, the cosmological bound can be relaxed if the gravitino 
is heavier than several tens TeV \cite{Jeong:2011sg}
and if the R-parity is broken by a small amount.

There exists yet another serious cosmological obstacle. 
To see this, first let us note that both real and imaginary components of the moduli fields
acquire the same mass, if they are stabilized by a large SUSY mass. This is the case if the moduli are stabilized
by the fluxes or by the KKLT mechanism. 
On the other hand, some of the moduli may be stabilized by SUSY breaking effects such that their
axionic fields remain extremely light due to the shift symmetry (\ref{shiftsym}). 
Indeed, in many string models, there are
often such ultralight axions~\cite{Svrcek:2006yi,Conlon:2006tq,Choi:2006za,Arvanitaki:2009fg,Acharya:2010zx,Higaki:2011me,Cicoli:2012sz,Cicoli:2013rwa}.
Those axions remain massless unless the shift symmetry is broken by an appropriate non-perturbative effects generated
in the low energy.
Furthermore, their real component partners tend to be lighter than those stabilized {\it a la} KKLT, 
because they are stabilized through the SUSY-breaking effect and
their masses are comparable to or lighter than the gravitino mass.
Therefore, it is crucial to study the cosmological impact of such light moduli stabilized by SUSY breaking effects,
as the lightest moduli fields usually play the most important role in cosmology.

In this paper we will show that such modulus generally decays into a pair of axions, contributing to 
dark radiation whose abundance is tightly constrained
by the recent Planck data~\cite{Ade:2013lta}. We call this problem {\it the moduli-induced axion problem}.
As we shall see shortly, this constrains a large portion of the parameter space, and most importantly, the problem persists even for a modulus mass
heavier than several tens TeV. The presence of such light moduli and ultralight axions may be a natural outcome of the string theories,
although it certainly depends on the details of the model such as the properties of compact geometry and brane configurations.
Indeed, if the strong CP problem is solved by the string theoretic QCD axion~\cite{Conlon:2006tq,Choi:2006za,Cicoli:2012sz}, 
it implies that there is at least one such modulus. 
Therefore, we believe that the moduli-induced axion problem is universal, and its solutions will provide us
with important information on the high energy physics. 

Lastly let us mention the related works in the past. 
It was recently pointed out in Ref.~\cite{Higaki:2012ba}
that the modulus decays into a pair of its axions in a context of the (moderately) 
LARGE volume scenario (LVS) \cite{Balasubramanian:2005zx},
and the produced axions will behave as extra radiation since the axions are effectively massless. 
Furthermore, the decay process has been extensively studied in Refs.~\cite{Cicoli:2012aq,Higaki:2012ar}, 
focusing on a possibility that the produced axions explains the excess of dark radiation hinted 
by the observation at that time. In the context of the SUSY QCD axion, it is well known that the QCD saxion
tends to decay mainly into a pair of QCD axions~\cite{Chun:1995hc}. 
The abundance of relativistic axions produced by the saxion 
decays was studied in Refs.~\cite{Choi:1996vz,Chun:2000jr,Ichikawa:2007jv,Jeong:2012hp,Choi:2012zna,Graf:2012hb,Bae:2013qr,Jeong:2013axf}.
In particular, several ways to suppress the branching fraction of the saxion decaying into axions were
discussed in Ref.~\cite{Jeong:2013axf}.

The rest of this paper is organized as follows. 
In Sec.~\ref{sec:2} we evaluate the modulus decay rate into axions and the Standard Model (SM) particles in a general setup
and show the robustness of the moduli-induced axion problem.
Possible solutions to the problem will also be mentioned.
In Sec.~\ref{sec:3}, we study concrete supergravity setups motivated by the string theory in order to
illustrate the moduli-induced axion problem.
Sec.~\ref{sec:4} is devoted to conclusions and discussion.

\section{Moduli-induced axion problem}
\label{sec:2}

Let us start with a simple low-energy effective theory containing one light modulus
stabilized by SUSY breaking effects. We consider the following K\"ahler potential,
\beq
K(T, T^\dag) \;=\; K(T+T^\dag),
\eeq
which respects the shift symmetry (\ref{shiftsym}).
Here and in what follows, we call
the real component of $T$ the modulus, whereas the axion refers to its imaginary component.
For the moment we assume that the modulus is stabilized by the K\"ahler potential and it does not
appear in the superpotential so that axion remains massless. 
Later we consider a case where the axion has a small but non-zero mass. 
The relevant terms in the Lagrangian are summarized in Appendix~\ref{sec:int}\footnote{See also \cite{Higaki:2011bz}.}.
In the following we adopt the Planck unit in which $M_P \simeq 2.4 \times \GEV{18}$ is set to be unity. 
We also denote the lowest component of a superfield by the same letter.

Let us define
\beq
T-\la T \ra \;\equiv\; \frac{1}{\sqrt{K_{TT}}} \frac{\tau + ia}{\sqrt{2}}.
\eeq
where $\tau$ and $a$ are (canonically normalized) real and imaginary components of $T$, respectively.
Here and in what follows, the subscript $T$ denotes the partial derivative with respect to $T$.
The partial decay rate into a pair of axions is
given by
\beq
\Gamma_a \;\equiv\; \Gamma(\tau \rightarrow aa) = \frac{1}{64 \pi} \frac{K_{TTT}^2}{K_{TT}^3} m_\tau^3,
\label{axdecay}
\eeq
where $m_\tau$ is the mass of $\tau$.
The modulus also couples to the axino, $\tilde a$, the fermionic partner of the axion.
The partial decay rate of the modulus into the axino pair is given by
\begin{equation}
	\Gamma(\tau\to \tilde a\tilde a) = \frac{1}{8\pi}\frac{K_{TTT}^2}{K_{TT}^{3} } m_{\tilde a}^2 m_\tau 
	\left(1- \frac{4m_{\tilde a}^2}{m_\tau^2} \right)^{3/2},
	\label{axinodecay}
\end{equation}
where $m_{\tilde a}$ denotes the axino mass. (See Appendix~\ref{sec:int}.)
The rate (\ref{axinodecay}) is suppressed by a factor of $\sim (m_{\tilde a}/m_\tau)^2$ 
with respect to (\ref{axdecay}).

In general, the modulus $\tau$  decays also into matter fields.
Suppose that the modulus has the following coupling,
\beq
K \;\supset\; Z_u |H_u|^2 + Z_d |H_d|^2 + g(T+T^\dag)\left(H_u H_d + {\rm h.c.}\right),
\label{K_THH}
\eeq
where $H_u$ and $H_d$ are up- and down-type Higgs superfields, respectively,
 $Z_{u(d)}$ is a K\"ahler metric for $H_{u(d)}$ that depends on the moduli, and $g$ is some function of $T+T^\dag$.
The partial decay rate
of $\tau$ into Higgs bosons is\footnote{
 If the decay into heavy Higgs is kinematically forbidden and the modulus decays only into
$hh$, $ZZ$ and $WW$,  the rate should be multiplied with $(\sin^2 2\beta)/2$, where $\tan \beta \equiv \la H_u^0 \ra/\la H_d^0 \ra$.
}
\beq
\Gamma(\tau \rightarrow HH) \; \simeq \; \frac{1}{8 \pi} \frac{g_T^{2}}{K_{TT} Z_u Z_d} m_\tau^3,
\label{tauHH}
\eeq
where we have neglected the mass of the Higgs bosons. 
The decay rate into higgsinos is given by
\begin{equation}
	\Gamma(\tau\to \tilde H\tilde H) = \frac{1}{8\pi}\frac{|c_{\tau \tilde h\tilde h}|^2}{K_{TT} Z_uZ_d }m_\tau,
	\label{tauhiggsino}
\end{equation}
where 
\begin{equation}
	c_{\tau \tilde h\tilde h} 
	= (2g_T+gK_T)m_{3/2}+ g_T(F^{T*}_T+F^{T*}_{\bar T}) + 2g_{TT}F^{T*}
	-2\left(\frac{\partial_T Z_u}{Z_u}+ \frac{\partial_T Z_d}{Z_d} \right)(gm_{3/2}+g_T F^{T*}).
\end{equation}
Here, $m_{3/2}$ denotes the gravitino mass and $F^T = -e^{K/2}K^{T\bar j}(D_jW)^*$ is the modulus $F$-term.

Similarly, if $T$ contributes to the SM gauge kinetic function $f_{\rm vis}$,
the modulus can decay into gauge bosons with the rate,
\beq
\Gamma(\tau \rightarrow A_{\mu}A_{\mu}) \;=\; \frac{N_g}{128\pi} \frac{|\partial_T f_{\rm vis}|^2}{({\rm Re} f_{\rm vis})^2} \frac{m_\tau^3}{K_{TT}},
\label{tauAA}
\eeq
where $N_g$ represents the number of gauge bosons, and it is given by $N_g = 8, 3, 1$ for $SU(3), SU(2)$ and $U(1)$, respectively. 
The decay into gauge bosons is sizable if $|\partial_T f_{\rm vis}|/({\rm Re} f_{\rm vis})$ is of order unity.
The decay rate into gauginos $(\lambda)$ is given by
\beq
\Gamma(\tau \rightarrow \lambda \lambda) \;=\; \frac{N_g}{128\pi}  \frac{|(\partial_T f_{\rm vis})(F^T_T+F^T_{\bar T}) +(\partial^2_T f_{\rm vis})F^T-2(\partial_Tf_{\rm vis}) m_\lambda|^2}{({\rm Re} f_{\rm vis})^2}
\frac{m_\tau}{K_{TT}},
\label{taugg}
\eeq
where $m_\lambda$ denotes the gaugino mass.
It also depends on the modulus F-term and it is (at most) comparable to that into gauge bosons. 
In particular, for a generic K\"ahler potential, the partial decay rate into gauginos is not suppressed by the gaugino mass~\cite{Endo:2006ix}.

For a generic K\"ahler potential the decay rate into axions, $\Gamma(\tau \rightarrow aa)$, is comparable to what is expected based
on the dimensional argument. 
Therefore there is no {\it a priori} reason to expect that the decay into axions 
is negligibly small with respect to the other decay processes, and so,  the branching fraction is generically
sizable, $B_a \equiv {\rm Br}(\tau \rightarrow aa) = {\cal O}(0.1)$.

In order to get the feeling that the branching fraction of the axion production tends to be
large, let us consider a simple example before continuing further.  In the next section we will study a few examples based on more 
realistic moduli stabilization. Consider the following K\"ahler potential of the no-scale form,
\bea
K&=&-3 \log\left[T+T^\dag - \frac{1}{3} \left\{|H_u|^2+|H_d|^2 + \left( z H_u H_d + {\rm h.c.} \right) \right\}\right] + \cdots,\\
 &=& -3 \log\left(T+T^\dag\right)+ \frac{1}{T+T^\dag} \left( |H_u|^2+|H_d|^2 + \left( z H_u H_d + {\rm h.c.} \right)\right) + \cdots,
\eea
where we assume that the modulus $\tau$ is stabilized by higher order terms not shown here, and $z$ denotes a coupling constant.
The superpotential and the gauge kinetic function are assumed to be
irrelevant for the modulus decay. 
The decay into the higgsino pair vanishes in the no-scale model as can be easily checked by using 
 (\ref{tauhiggsino}).
In the no-scale model $T$ dominantly breaks SUSY and $\tilde a$ becomes goldstino `eaten' by  gravitino.
Since $m_\tau < m_{3/2}$ in the no-scale model, the decay into a pair of $\tilde a$ (or the gravitino) is kinematically forbidden.
Then the branching fraction of the axion production is given by
\bea
B_a &=& \frac{1}{2z^2+1}.
\label{ba}
\eea
Therefore, $B_a$ is indeed of order $0.1$ for $z = {\cal O}(1)$.

As mentioned in the Introduction, the presence of additional relativistic degrees of freedom is tightly constrained
by the Planck results. The constraint on the effective number of neutrinos, $N_{\rm eff}$,  reads~\cite{Ade:2013lta}
\beq
N_{\rm eff} = 3.30^{+0.54}_{-0.51} ~~~(95\%;~{\rm Planck + WP + highL + BAO}).
\label{NeffPl}
\eeq
In the present scenario, $\Delta N_{\rm eff}(\equiv N_{\rm eff} - 3.046)$ is related to $B_a$ as
\bea
\Delta N_{\rm eff} 
&=&  \frac{43}{7} \lrfp{10.75}{g_*(T_d)}{\frac{1}{3}} \frac{B_a}{1-B_a},
\eea
where $g_*$ counts the relativistic degrees of freedom at $T=T_d$, and $T_d$ is the decay temperature of the modulus
defined by
\beq
T_d \;=\; (1-B_a)^\frac{1}{4} \lrfp{\pi^2 g_*}{90}{-\frac{1}{4}} \sqrt{\Gamma_{\rm total} M_P},
\eeq
with $\Gamma_{\rm total}$ being the total decay rate of the modulus when the modulus dominates over the energy density of the Universe.
Thus, the branching fraction is bounded above as $B_a \lesssim 0.12 \sim 0.22$ for $g_* = 10.75 \sim 106.75$.

\begin{figure}[t!]
\begin{center}
\includegraphics[width=10cm]{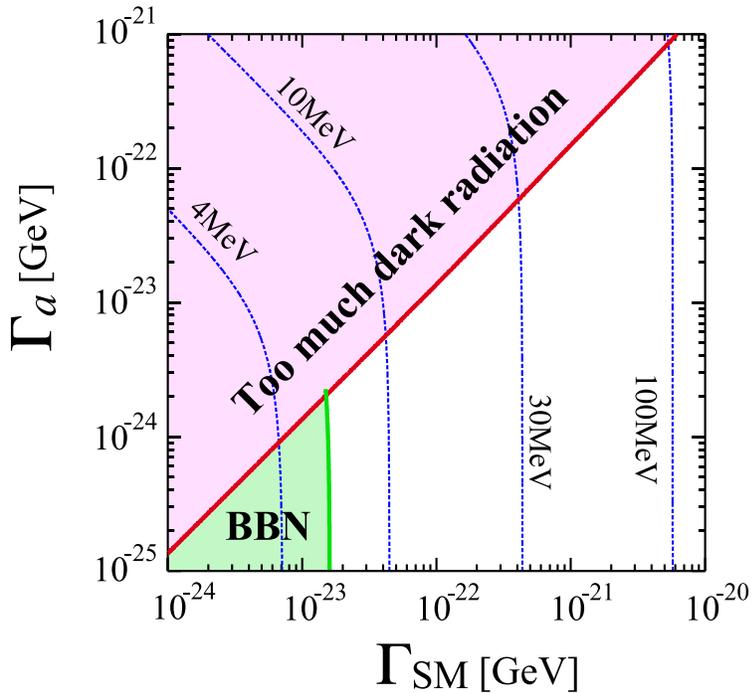} 
\caption{The cosmological bounds on the partial decay rates, $\Gamma_a$ and $\Gamma_{SM}$, are shown.
In the upper left shaded (pink) region, the axionic dark radiation is overproduced, leading to $\Delta N_{\rm eff} > 0.84$.
In the lower left shaded (green) region, the modulus decay temperature is lower than $6$\,MeV, and
the $^4$He abundance is too large to be consistent with observations~\cite{Kawasaki:1999na}. 
The dotted (blue) lines are contours of the decay temperature, $T_d = 4, 10, 30, 100$\,MeV from left to right.
In this figure, we have not considered the LSP overproduction through the decay.
}
\label{fig:const}
\end{center}
\end{figure}

We show the cosmological constraints on the partial decay rates  $\Gamma_{a}$ and $\Gamma_{\rm SM}$
in Fig.~\ref{fig:const}, where $\Gamma_{\rm SM}$ denotes the  decay rate into the SM particles. 
In the lower left shaded (green) region, the modulus decay temperature is lower than $6$\,MeV, and
the $^4$He abundance is too large to be consistent with observations~\cite{Kawasaki:1999na}.\footnote{
Note that our definition of $T_d$ is higher than that defined in \cite{Kawasaki:1999na} by a factor of $\sqrt{3}$.
}  In the upper left shaded (pink) region, the axionic dark radiation is overproduced, leading to $\Delta N_{\rm eff} > 0.84$.
It is worth noting that the dark radiation constraint extends to the decay temperature much higher 
than $6$\,MeV. In other words, the moduli-induced axion problem is not solved by simply increasing the modulus mass.
Note also that the constraint can be much more severe 
if the axion has a small but non-zero mass and decays into photons, 
electrons, etc. at late time.

We have here neglected the modulus decay into higgsinos and gauginos
because the rates depend on the modulus stabilization mechanism and the 
relation between the modulus and higgsino/gaugino masses. Even if these decay processes
are taken into consideration, the constraints shown in Fig.~\ref{fig:const} will not be changed much.
In this respect, the moduli-induced axion problem is robust.
 Note that the abundance of the LSPs produced by the modulus decay may exceed the observed dark matter
abundance, if $T_d \lesssim 1$\,GeV and R-parity is conserved. This is especially the case if
the decay rate into higgsinos and/or gauginos is comparable to that into Higgs and/or gauge bosons. 
The LSP overproduction can be avoided if the LSP annihilation cross section is relatively large or the R-parity is violated 
by a small amount.

Lastly let us discuss possible solutions to the moduli-induced axion problem. There are basically two ways to solve the  problem.
One is to increase the decay rate into the SM particles by introducing additional decay channels or setting the coupling
constants larger. For instance, in the aforementioned example, the branching fraction $B_a$ satisfies the Planck constraint
if $z \gtrsim 2$ (see Eq.~(\ref{ba})). Alternatively, we may introduce more than four pairs of Higgs doublet with a similar
coupling to the modulus with $z \simeq 1$.
The decay rate into SM gauge bosons is also sizable if the gauge kinetic function is modulus dependent. 
Actually, if the SM branes are wrapping on the cycle in stringy compactifications, 
the corresponding moduli couple to the gauge bosons through gauge kinetic functions.
An well-motivated example is the string theoretic QCD saxion, as we will see in Sec.~\ref{sec:3B}.
The other solution is to introduce an approximate $Z_2$ symmetry on $T$, which
suppresses $K_{TTT}/K_{TT}$. The $Z_2$ symmetry can be just a coincidence because one order of magnitude suppression 
is sufficient to satisfy the bound. However, one needs to make sure that the decay rate into the SM particles should not be
similarly suppressed. In the next section we will take up two examples based on concrete moduli stabilization to illustrate
the moduli-induced axion problem and its solutions.

\section{Examples}
\label{sec:3}

In this section, we shall consider two examples, where
an axion will appear at low energy scales: LVS \cite{Balasubramanian:2005zx} 
and KKLT-like one \cite{Choi:2006za}.
In the LVS case, the axion remains light  
because large extra dimensions suppress any shift-symmetry breaking effects,
while in the latter case there exists a specific geometry which preserves the shift symmetry. 
One will find that the production of axion dark radiation is unavoidable in the LVS once the lightest modulus dominates 
over the energy density of the Universe. On the other hand, in the latter case, 
the decay rate into axions can be suppressed by choosing an appropriate internal geometry,
which implies a mild fine-tuning between the intersection number among K\"ahler forms on a CY space (K\"ahler potential) 
and such a geometry (superpotential).

\subsection{LARGE volume scenario}

Let us consider an effective action of K\"ahler moduli on a Swiss cheese CY space in flux vacua\footnote{
Depending on orientifolding and D-brane configuration, moduli is renormalized at 1-loop level \cite{Conlon:2009kt}.
Then soft mass on visible brane mentioned below will be changed via such a quantum effect \cite{Conlon:2010ji}.
In such cases, we will still have dark radiation via the 
lightest modulus decay \cite{Higaki:2012ar}.
} \cite{Balasubramanian:2005zx}:
\begin{align}
K_{\rm moduli} &=  - 2\log\bigg({\cal V} + \frac{\hat{\xi}}{2} \bigg) ; \qquad 
{\cal V} = (T_b + T_b^{\dag})^{3/2}- (T_s + T_s^{\dag})^{3/2}, \\
W_{\rm moduli} &= W_{0} + A e^{-a_s T_s} ; \qquad a_s = \frac{2\pi}{N} .
\end{align}
Here, $T_b$ and $T_s$ are the K\"ahler moduli and ${\cal V}$ is the CY volume; 
$\hat{\xi}$, $W_0$ and $A$ are of ${\cal O}(1)$ constants. $N$ is a positive integer.
We have assumed that the 4-cycle $S$ supported by ${\rm Re}(T_s)$ is rigid divisor, i.e. $h^{1,0}(S)=h^{2,0}(S)=0$ 
\cite{Witten:1996bn}:
A non-perturbative effect can exist on such a divisor.
At the minimum of the scalar potential 
$V= e^{K}\left[|DW|^2-3|W|^2 \right] + \frac{1}{2}D^2$, where $DW \equiv (\partial K) W + \partial W$, one finds
\begin{align}
{\cal V} &\sim \frac{W_0}{a_s A} \exp(\hat{\xi}^{2/3}) \gg 1,
\qquad
a_s T_s  \sim  \hat{\xi}^{2/3}; \\
m_{a} &= 0, \quad m_{\tau_b} \sim \frac{1}{{\cal V}^{3/2}} , \quad
m_{3/2} \sim \frac{1}{{\cal V}}, \quad
m_{T_s} \sim \frac{\log({\cal V})}{{\cal V}}.
\label{LVS}
\end{align}
with broken SUSY\footnote{The vacuum has a negative cosmological constant $\sim - m_{3/2}^3M_{P}$,
however, we will not consider a concrete uplifting potential. Including the uplifting potential does not drastically 
change the results in this subsection because of the small negative cosmological constant owing to the SUSY breaking. 
This should be contrasted to the KKLT case.}.
Here $a$ and $\tau_b$ are the canonically normalized ${\rm Im}(T_b)$ and ${\rm Re}(T_b)$ respectively, 
and $m_{3/2} = e^{K/2}W$ the gravitino mass; the lightest modulino is the goldstino.  
Thus we can clearly see that the overall volume modulus, $\tau_b$ is the lightest modulus and it may have dangerous cosmological effects. 
The axion $a$ stays ultralight owing to the large volume CY,
even if $e^{-2\pi T_b} \propto e^{2\pi {\cal V}^{2/3}}$ is included in the scalar potential. 
Hence axions produced by the modulus decay behave as dark radiation \cite{Cicoli:2012aq,Higaki:2012ar}.
By using (\ref{axdecay}), the modulus partial decay width into the axion pair is calculated as
\begin{equation}
	\Gamma(\tau_b \to aa) = \frac{1}{48\pi}\frac{m_{\tau_b}^3}{M_P^2}.
\end{equation}
On dimensional grounds, this is an expected order of magnitude, 
which,  therefore, cannot be significantly suppressed compared to the other decay processes. 
Thus there is the moduli-induced axion problem, which persists even for heavy modulus mass.
In this model, the fermionic superpartner of the modulus is the goldstino and obtains a mass of $m_{3/2}$.
Hence the modulus decay into its fermionic component is kinematically forbidden. 
To see how severe the moduli-induced axion problem is, let us consider the modulus interaction with the SM sector.

Suppose that there is a singular cycle moduli denoted by $T_v$ which supports the SM branes\footnote{
We can have another odd-parity moduli under the orientifold parity, 
however it is irrelevant for us because it will be also stabilized via another anomalous $U(1)$ D-term, similarly to $T_v$
\cite{Conlon:2008wa}.
}:
\begin{align}
K_{\rm sin} &=  \frac{(T_v +T_v^{\dag} +V_{U(1)})^2}{{\cal V}} + K_{\rm matter}(Q,Q^{\dag}),~~
W_{\rm sin} = W_{\rm matter}(Q) , ~~ f_{\rm sin} = \frac{T_v}{4\pi} + f_0 , 
\end{align}
\begin{align}
K_{\rm matter}(Q,Q^{\dag}) &= e^{K_{\rm moduli}/3} [ |Q|^2 + (z H_u H_d + {\rm c.c.})] .
\end{align}
Here $Q$ denotes all visible matter superfields including Higgs ones,
$f_0$ is a constant and we have neglected higher order terms of $T_v$ in $K_{\rm matter}$.
$V_{U(1)}$ is the anomalous $U(1)$ gauge multiplet, and $z$ is the coupling constant.
If Higgs sector has a non-chiral origin like in a case of Gauge-Higgs Unification, $z \simeq 1$ is expected
\cite{Hebecker:2012qp};
we will consider the case of $z ={\cal O}(1)$.
In addition to (\ref{LVS}), we find
\begin{equation}
	T_v  = 0, ~~~ m_{T_v} \sim \frac{1}{{\cal V}^{1/2}}.
\end{equation}
Note that $T_v$ is absorbed into $V_{U(1)}$.
In this setup, by using (\ref{tauHH}), we obtain the modulus partial decay width to the Higgs boson pair as
\begin{equation}
	\Gamma(\tau_b \to HH) = \frac{z^2}{24\pi}\frac{m_{\tau_b}^3}{M_P^2}.
\end{equation}
The decay into the higgsino pair is much suppressed due to the approximate no-scale structure of the K\"ahler potential,
as mentioned in Sec.~\ref{sec:2}.
Thus the branching fraction into the axion pair is given by $B_a \simeq 1/(1+2z^2)$, which is same as the previous 
estimate (\ref{ba})  because the effective action for $T_b$ possesses a no-scale structure up to a correction of $1/{\cal V} \ll 1$.
Therefore, we need $z \gtrsim 2$ or more than four pairs of Higgs doublet
to avoid the axion overproduction from the modulus decay.

Let us comment on the soft masses in this model.
On the visible branes, the SUSY-breaking soft mass is given by
\begin{align}
m_{\rm soft} \sim F^{\rm local} \sim \frac{1}{{\cal V}^2} < m_{\tau_b}.
\end{align}
Here, $m_{\rm soft}$ includes higgsino mass and $B$-term,
and $F^{\rm local}$ denotes the $F$-components of the local modulus, i.e., dilaton.
This is because the SM sector is a local model decoupled from the bulk, and 
the dilaton (and complex structure moduli) are stabilized supersymmetrically through a flux compactification,
then affected by the $\alpha$'-correction $\hat{\xi}/{\cal V}$ in the K\"ahler potential; 
for ${\cal V} = {\cal O}(10^7)$, one finds $m_{\rm soft} ={\cal O} (1-10)$ TeV and $m_{\tau_b} ={\cal O}(10^7)$ GeV.\footnote{
We expect that $\tau_b$ is coupled to the SM gauge bosons through a quantum effect: 
$\frac{\alpha_{\rm SM}}{4\pi}\frac{\tau_b}{M_{P}}(F_{\mu \nu})^2$ \cite{Dixon:1990pc}.
}
The decay temperature of the modulus is given by
\begin{align}
T_d =
1.7\, {\rm GeV} \times  (1-B_a)^\frac{1}{4} \left(\frac{2z^2+1}{3}\right)^{1/2}
\left(\frac{80}{g_*(T_d)}\right)^{1/4} 
\left(\frac{m_{\tau_b}}{10^7{\rm GeV}}\right)^{3/2}.
\end{align}
If the neutral Wino is the lightest supersymmetric particle (LSP) with the conserved R-parity,
it will become a good candidate of cold dark matter
because the modulus decay can produce abundant Winos at $m_{\tilde{W}} \sim 700$GeV~\cite{Higaki:2012ar}.

\subsection{KKLT: A string-inspired QCD axion model}
\label{sec:3B}

Let us again consider a similar effective action on a CY space in flux vacua
as a generalization of the setup considered in Ref.~\cite{Conlon:2006tq,Choi:2006za}:
\begin{align}
K_{\rm moduli} &=  - 2\log ( {\cal V}  ) ; \qquad 
{\cal V} = (T_0 + T_0^{\dag})^{3/2}- \kappa_1 (T_1 + T_1^{\dag})^{3/2} - \kappa_2 (T_2 + T_2^{\dag})^{3/2}, \\
W_{\rm moduli} &= W_{0} + A e^{-\alpha T_0} + B e^{-\beta(T_1 + n T_2)} ; \qquad \alpha = \frac{2\pi}{N},~~~\beta= \frac{2\pi}{M} .
\end{align}
Here $W_0 \ll 1$ via a fine-tuning of the fluxes, $N,~M$ are positive integers, and $A\sim B \sim \mathcal O(1)$.
$\kappa_{1,2} > 0$ are constants which depend on intersection numbers among K\"ahler forms (two cycles) on the CY space,
and $n \in {\mathbb Z}$ depends on the configuration of an Euclidean (instanton) brane.
Note that it is only $\Phi \equiv T_1 + n T_2$ that is stabilized in a SUSY way,
and the axion multiplet $\mathcal A \equiv n T_1 - T_2$ is absent in the superpotential
because we have assumed that only the two 4-cycles supported by ${\rm Re}(T_0)$ and ${\rm Re}(T_1 + n T_2)$ 
are rigid divisors. Hence one could not find any other non-perturbative effects\footnote{
For instance, flux can affect the zero mode spectra. However, we will not consider such a possibility for  simplicity.
}.
As we will see, the imaginary component of $\mathcal A$ can play a role of the QCD axion to solve the strong CP problem.

At the minimum of the scalar potential, adding a sequestered uplifting potential $\delta V = \epsilon e^{2K_{\rm moduli}/3}$
to realize the Minkowski vacuum through a fine-tuning\footnote{
For a non-sequestered potential $\delta V_{\rm non-seq} = \epsilon e^{K_{\rm moduli}}$, we will have similar results \cite{Acharya:2010zx,Higaki:2011me}.
},
one obtains solutions near the supersymmetric location $D_{T_0}W \simeq D_{\Phi}W \simeq \partial_{\cal A}K \simeq 0$:
\begin{align}
\alpha T_0 &\simeq \beta \Phi \simeq \log \left(\frac{1}{W_0}\right), \qquad
{\rm Re}(\mathcal A) \simeq n {\rm Re}(\Phi) \bigg(\frac{-n\kappa_1^2 +\kappa_2^2}{n^3 \kappa_1^2 + \kappa_2^2}\bigg) ; \\
m_a = 0, \qquad m_{3/2} &\simeq W_0, \qquad
m_{s} \simeq \sqrt{2}m_{3/2}, \qquad  m_{T_0} \simeq m_{\Phi} \simeq \log\left(\frac{M_P}{m_{3/2}}\right)m_{3/2}.
\end{align}
Here, $a$ and $s$ are the canonically normalized ${\rm Im}(\mathcal A)$ and ${\rm Re}(\mathcal A)$ respectively.
We will assume that ${\cal V} > 0$ is obtained at this minimum, i.e., $\alpha < \beta$.
The mixing between the saxion and the other moduli is suppressed by a power of $1/\log(M_{P}/m_{3/2})$, 
hence we will ignore such a mixing.
Hereafter we will focus on the cosmological effects of $s$ since it is the lightest modulus.
Note that axino mass is given by $m_{3/2}$ whereas the modulino masses are 
same as the corresponding moduli masses.
Let us estimate the partial decay width of saxion $s$ into the axion pair.
By using (\ref{axdecay}), one finds
\begin{align}
\Gamma_a &\simeq 
\frac{ \left(n^3 {\kappa_1}^2 - {\kappa_2}^2\right)^2 }
{768  \pi  {\kappa_2 }^3 }
\frac{M_S^3}{M_P^2}, 
\end{align}
where
\begin{align}
M_S^3 &\equiv \frac{1}{\sqrt{2}}\bigg(\frac{\sqrt{n^3 {\kappa_1}^2+ {\kappa_2}^2}}{n^3 {\kappa_1}^3}\bigg)
\frac{{\cal V}}{\phi^{3/2}} m_s^3 .
\end{align}
Here $\phi \equiv {\rm Re}(\Phi)$.
Hence, the saxion partial decay width into the axions is suppressed if
\begin{align}
n^3  \simeq \bigg(\frac{\kappa_2}{\kappa_1}\bigg)^2.
\label{tune1}
\end{align}
For $n=1$, the axion overproduction is significantly relaxed if the volume of the 4-cycles characterized by $T_1$ and $T_2$ are  symmetric under exchange.

Now let us see the modulus coupling to the SM sector.
We will assume that the SM localizes on the D-branes wrapping on the cycle supported by $T_2$ \cite{Conlon:2006tj}:
\begin{align}
K_{\rm matter}(Q,Q^{\dag}) &=  \frac{1}{{\cal V}^{2/3}}\bigg[ (T_2 +T_2^{\dag})^{\lambda} |Q|^2 + 
(T_2 +T_2^{\dag})^{\lambda_{\rm GM}}(z H_u H_d + {\rm c.c.})\bigg] 
, \\
W_{\rm vis} &= W_{\rm matter}(Q) , ~~~ f_{\rm vis} = \frac{T_2}{4\pi} = \frac{1}{4\pi}\frac{n\Phi - \mathcal A}{n^2 +1}.
\end{align}
Here $0 \leq \lambda \leq 1$. Assuming that the Higgs sector is non-chiral, 
one will obtain $\lambda_{\rm GM} = \lambda$ and $z \simeq 1$. 
We will take the minimal case of $\lambda = \lambda_{\rm GM} = 1/3$ for simplicity.
In this setup, the imaginary component of $\mathcal A$ behaves as the QCD axion,
since it is massless at the perturbative level and obtains a potential dominantly from the QCD instanton effect. 
The decay constant of the QCD axion is given by
\begin{align}
f_a = (n^2+1)\frac{\sqrt{2 K_{\mathcal A \mathcal A}}}{2\pi} \simeq \frac{1}{2 \pi {\cal V}^{1/2}\phi^{1/4}} .
\end{align}
As for the soft mass on the visible brane, we will obtain mass spectra through the mirage mediation~\cite{Choi:2004sx,Endo:2005uy}:
\begin{align}
m_{\rm soft} \simeq \frac{F^{T_2}}{T_2 +T_2^{\dag}} \simeq 
\frac{m_{3/2}}{\log\left(\frac{M_P}{m_{3/2}}\right)} < m_s .
\end{align}
If there is the Giudice-Masiero term with $z \simeq 1$, one finds $|\mu|^2 \simeq B\mu \simeq m_{3/2}^2$\footnote{
In the MSSM, such large (soft) terms in the Higgs sector is dangerous because a color-charge-breaking minimum 
can be realized then \cite{Casas:1995pd}.
In this paper, however, we will not consider such a problem seriously as the Higgs potential depends on the model.
}.
For $m_{3/2} ={\cal O}(10^6)$ GeV, one obtains $\log(M_P/m_{3/2}) \simeq 30$ and $m_{\rm soft} ={\cal O} (10)$TeV;
moduli VEVs will be of ${\cal O}(1-10)$, then $f_a ={\cal O}(10^{16})$ GeV.

The partial decay widths of saxion into the SM sector are estimated from (\ref{tauHH}) and (\ref{tauAA}) as
\begin{align}
\Gamma (s \to hh, WW,ZZ) &\simeq
z^2 \sin^2 (2\beta)\frac{  \lambda^2 {\kappa_2}  }
{48  \pi  } \frac{M_S^3}{M_P^2} , \qquad
\Gamma (s \to A_\mu A_\mu) \simeq
N_g \frac{  {\kappa_2} }
{96  \pi } \frac{M_S^3}{M_P^2} ; 
\end{align}
On the estimation of the decay fraction into Higgses, we have assumed that the saxion decay into the heavy Higgses is not allowed kinematically. 
The saxion decay rate into gauginos is estimated by using 
\begin{align}
\partial_{\mathcal A} F^{T_2} = \partial_{\bar{\mathcal A}} F^{T_2}\simeq \frac{1}{n^2 + 1} \frac{m_s}{\sqrt{2}},
\end{align}
which is led from the fact that $F^{T_i}/(T_i+T_i^{\dag})$ is universal for $^\forall i$.
Thus, a sizable contribution to that into gauge boson is found from (\ref{taugg}):
\begin{align}
\Gamma (s \to \lambda \lambda) \simeq N_g \frac{  {\kappa_2} }{48  \pi } \frac{M_S^3}{M_P^2} .
\end{align}
Note that higgsinos obtain masses comparable to the saxion for $z\sim 1$, hence decays into higgsinos can be forbidden.
Otherwise, it would give a comparable contribution to that into Higgs bosons.
Then the total decay width of the saxion and the decay temperature become
\begin{align}
\Gamma_{\rm total} & \simeq \frac{c}{48\pi}M_S^3; ~
c \equiv 
\frac{\left(n^3 {\kappa_1}^2 - {\kappa_2}^2\right)^2 }{16{\kappa_2 }^3}
+ 
z^2 \sin^2 (2\beta)\lambda^2 {\kappa_2}  + \frac{3\kappa_2 N_g }{2} , \\
T_d & \simeq 0.20\,{\rm GeV} \times (1-B_a)^{1/4}\left(\frac{c}{30}\right)^{1/2}\left(\frac{40}{g_*(T_d)}\right)^{1/4} \left(\frac{M_S}{10^{6}\,{\rm GeV}}\right)^{3/2} ,\\
B_a &= \frac{\left(n^3 {\kappa_1}^2 - {\kappa_2}^2\right)^2}{\left(n^3 {\kappa_1}^2 - {\kappa_2}^2\right)^2
+ 16\kappa_2^4z^2 \sin^2 (2\beta)\lambda^2 +24\kappa_2^4 N_g } .
\end{align}

\begin{figure}[t!]
\begin{center}
\includegraphics[width=11cm]{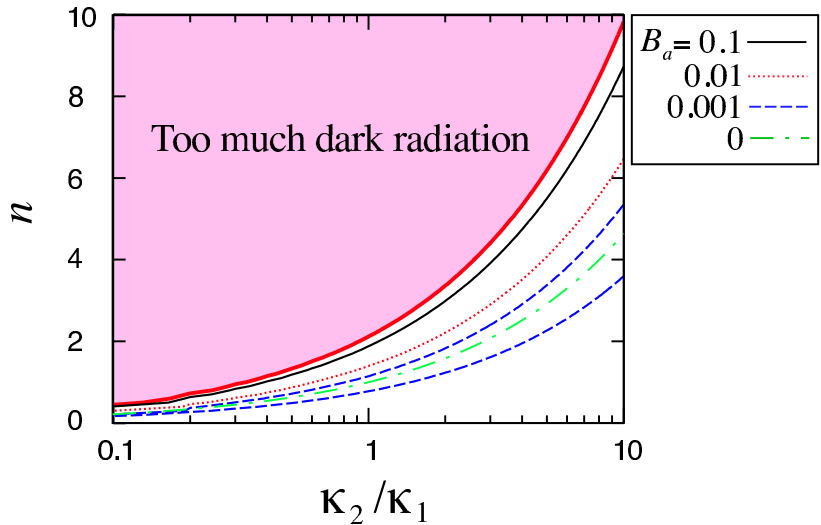} 
\caption{Contours of the saxion branching fraction into the axion pair $B_a$ on ($\kappa_2/\kappa_1, n$) plane.
In this plot we have taken $z=0$ and $N_g=12$.
The shaded region is excluded from the axion overproduction.
}
\label{fig:Ba}
\end{center}
\end{figure}

There are several ways to suppress the branching ratio into axions.
Obviously, if the condition (\ref{tune1}) is satisfied, the decay rate into axions is suppressed.
Alternatively, the increase of the decay width into Higgs suppresses $B_a$, which is
realized  by taking $z \gtrsim 1$ for unsuppressed $\kappa_2$.
Similar suppression is obtained  if there are more than two pairs of Higgs doublet.
Moreover, for $\kappa_2 \gg n\kappa_1$ the branching fraction becomes suppressed due to the relative enhancement 
of the decay into gauge bosons: $B_a \simeq 1/(24N_g)$.
These features are seen in Fig.~\ref{fig:Ba} where we plotted the contours of $B_a$ on ($\kappa_2/\kappa_1, n$) plane.
In this plot we have taken $z=0$ and $N_g=12$.
The shaded region is excluded from the axion overproduction.

Note that, since $m_{\rm soft} \ll m_{3/2}$, the LSPs are likely  overproduced by the saxion decay into gauginos~\cite{Endo:2006ix}.
To avoid the LSP overproduction, the R-parity should be broken by a small amount, and the QCD axion discussed in this section will be a candidate of dark matter, if a mild tuning of the initial misalignment angle is allowed.

For $n=0$, $T_2$ is not stabilized at this level, and hence quantum corrections on the K\"ahler potential 
and thus SUSY-breaking will stabilize it
\cite{vonGersdorff:2005bf,Berg:2004ek,Cicoli:2012sz}, 
giving the mass only to the saxion ${\rm Re}(T_2)$: $m_{{\rm Re}(T_2)} \lesssim m_{3/2}$ whereas $m_{{\rm Im}(T_2)} = 0$.
Thus the saxion will mainly decay into axions and the SM gauge bosons; as for the dark radiation 
the result will be similar to the case for $n>0$. 
Whether LSPs can be produced through the decay or not depends on the quantum corrections.
If the saxion is lighter than the SM gauginos it is not necessary for R-parity to be broken, 
while dark matter mainly consists of QCD axion ${\rm Im}(T_2)$ similarly.
Otherwise, R-parity should be violated.
For $n<0$, $K_{{\cal A}\bar{\cal A}}$ takes unphysical values.

\section{Discussion and Conclusions}
\label{sec:4}

In this paper, we have studied a cosmological fate of the lightest moduli appearing 
in string compactifications, assuming that the moduli dominate over the energy density 
of the Universe. In particular, we have focused on those moduli fields stabilized by SUSY breaking 
effects, as they tend to be lighter than those stabilized in a supersymmetric fashion. 
We have pointed out that those moduli fields can have an important cosmological
effect, even if they are so heavy that they decay before BBN.
This is because the moduli often dominantly decay into the  light axionic components, which would result in
too much axion dark radiation in contradiction with observations.
Considering that an imaginary component of one of the string moduli can play a role of the QCD axion to solve the strong CP problem,
it is plausible that there is at least one such modulus. 
Even without invoking the QCD axion, the axionic partner of the overall volume modulus is ultra-light in the LVS scenario.
Such a volume modulus decays into the axion with sizable branching fraction.
These examples nicely illustrate a new aspect of the cosmological moduli problem which cannot be solved just by setting
the modulus mass heavy.
We have called the axion overproduction problem from the modulus decay as the ``moduli-induced axion problem''.

If axions in a string vacuum obtain a small but non-zero mass, the constraint would become severer.
For instance, they might decay into radiation at late time if sufficiently heavy and unstable.
They will also contribute to the dark matter and its isocurvature perturbations, if stable.\footnote{
If there are multiple axions, one needs to follow their evolution in order of their masses,
as the isocurvature perturbations depend on their abundance and the size of quantum fluctuations.
For instance, this will be the case if multiple light axions appear in the SM gauge kinetic function, where
the QCD axion (mass eigenstate) receives quantum fluctuations of the axions with a possibly different
decay constant. 
}
To avoid these problems, we may need proper moduli stabilization for the former case, and
for the latter case, a fine-tuning of the initial misalignment angle or a low-scale inflation will be 
required additionally~\cite{Fox:2004kb,Acharya:2010zx}.

Now we are in a position to discuss possible solutions to the moduli-induced axion problem.
The branching fraction can be suppressed in two ways:
to increase the partial width into the SM sector such as Higgs fields and gauge bosons, or to reduce the partial width into the axion pair.
For example, the string theoretic QCD saxion can naturally have a sizable decay width into gauge bosons.
In this case, it is possible that there is a small amount of axion dark radiation, 
which may be detected in the future observations \cite{Conlon:2013isa}. 
On the other hand, the reduction of partial width into the axion pair may imply some approximate 
symmetry on the internal CY geometry.
As explicitly studied in Sec.~\ref{sec:3B}, if there is a symmetry $\kappa_1 \leftrightarrow \kappa_2$ for $n=1$,
the partial width into the axion pair is suppressed. It implies that the size of two 4-cycles are symmetric under  exchange.
In this case, it is also possible that the saxion is stabilized near the low-energy
minimum during inflation, and its oscillation amplitude can be significantly suppressed. 
If so, there may be negligible amount of axionic dark radiation in the present Universe.  
Of course, it is always possible to assume a huge amount of entropy production by some other (brane) fields 
which mainly decay into the SM particles. 

Depending on models, a heavy field, which is not the superpartner of an axion, 
also can produce the axion dark radiation \cite{Cicoli:2012aq,Conlon:2013isa}.
In this sense, this moduli-induced axion problem (and its solutions) lie not only in SUSY models but also non-SUSY models: 
For instance, the inflaton decay in (non-)SUSY models
can produce such axion dark radiation, if there exist axions in nature.

\section*{Acknowledgments}
We would be grateful to K.i. Okumura for useful discussion on the mirage mediation.
We would like to thank the YITP at Kyoto University for the hospitality
during the YITP workshop YITP-W-12-21 on ``LHC vs Beyond the Standard Model'',
where the present work started. 
This work was supported by 
the Grant-in-Aid for Scientific Research on Innovative Areas
(No.~24111702 [FT], No.~21111006 [KN and FT], and No.~23104008 [FT]),  
Scientific Research (A) (No.~22244030 [KN and FT] and No.~21244033 [FT]), 
JSPS Grant-in-Aid for Young Scientists (B) (No. 24740135 [FT] and No. 25800169 [TH]),
and Inoue Foundation for Science [FT].

\appendix

\section{Relevant moduli interactions}
\label{sec:int}

In this appendix we list relevant terms in the Lagrangian for the modulus decay.
We put several assumptions.
First, the modulus $T$ does not appear in the superpotential so that its axionic component remains massless.
We also assume that $T$ has negligible contribution to the SUSY breaking 
and negligible kinetic mixing between $T$ and the SUSY breaking field $z$: $K_{Tz^*}=0$.
In this Appendix, the canonically normalized fields are represented by hats and given by: 
$\hat T \equiv \sqrt{K_{TT}} T$, 
$\hat H_{u,d} \equiv \sqrt{Z_{u,d}}H_{u,d}$, 
$\hat A_\mu \equiv \sqrt{{\rm Re}f_{\rm vis}} A_\mu$.
$\hat \lambda \equiv \sqrt{{\rm Re}f_{\rm vis}} \lambda$.

\subsection{Moduli-axion}

From the axion kinetic term, we easily find
\begin{equation}
\begin{split}
	\mathcal L = -\frac{1}{\sqrt 2} K_{TTT}\tau(\partial a)^2 
	=  -\frac{1}{\sqrt 2} \frac{K_{TTT}}{K_{TT}^{3/2}} \hat \tau(\partial \hat a)^2 .
\end{split}
\end{equation}
This leads to the modulus partial decay width as (\ref{axdecay}).

\subsection{Moduli-axino}

From the axino kinetic term we soon find
\begin{equation}
\begin{split}
	\mathcal L_{\rm kin} =\frac{1}{\sqrt 2}K_{TTT} \tau \left[-i {\tilde a}^\dagger \bar\sigma_\mu\partial^\mu \tilde a 
	+ i (\partial_\mu {\tilde a}^\dagger)\bar\sigma^\mu \tilde a \right],
\label{tau_axino}
\end{split}
\end{equation}
which, by using the equation of motion of axino, becomes
\begin{equation}
	\mathcal L_{\rm kin} = \frac{1}{\sqrt 2}\frac{K_{TTT}}{K_{TT}^{3/2}} m_{\tilde a} 
	\hat \tau (\hat{\tilde a} \hat{\tilde a} + {\rm h.c.}),
\end{equation}
in terms of the canonically normalized fields. Here the the physical axino mass reads~\cite{Wess:1992cp}
\begin{equation}
	m_{\tilde a} = \frac{1}{K_{TT}}\langle e^{G/2}\rangle \langle \nabla_TG_T +\frac{1}{3}G_TG_T \rangle
	= m_{3/2}\left( 1 + \frac{K_T^2}{3K_{TT}} \right) + \frac{K_{TTT} F^*}{K_{TT}},
\end{equation}
where $G = K + \ln W + \ln W^*$, $\nabla_i G_j = \partial_i G_j - \Gamma^k_{ij} G_k$,
$\Gamma^k_{ij}=K^{k\bar l}K_{j\bar li}$,
and $F^T = - e^{K/2}K^{T\bar j}(D_jW)^*$ represents the size of modulus $F$-term.
Note that there are in general mixings between the axino and goldstino, but the mixing angle is suppressed by $\sim K_T$,
which is here assumed to be small.
We also assume ${K^{TT}}_T K_T \ll 1$.
Neglecting terms involving $K_T$, we find the modulus-axino interaction from the axino mass term as
\begin{equation}
	\mathcal L_{\rm mass} = -\frac{1}{2\sqrt 2 K_{TT}^{3/2}}\left[ K_{TTT}(2m_{3/2}+F^{T*}_T+F^{T*}_{\bar T})
	 \right] \hat\tau \hat{\tilde a}\hat{\tilde a} + {\rm h.c.}
\end{equation}
Noting that $F^{T*}_T= -m_{3/2}$ in the present approximation, the interaction $\mathcal L_{\rm mass}$ vanishes.
Therefore, the only relevant interaction is $\mathcal L_{\rm kin}$, from which we obtain the partial decay width as (\ref{axinodecay}).

\subsection{Moduli-Higgs}

The K\"ahler potential is given by Eq.~(\ref{K_THH}).
The interaction terms are given by
\begin{equation}
\begin{split}
	\mathcal L	=
	\frac{1}{\sqrt{2}} \left[ (\partial_T Z_u) \tau (H_u \partial^2 H_u^\dag +{\rm h.c.}) 
	+ (\partial_T Z_d) \tau (H_d \partial^2 H_d^\dag +{\rm h.c.})   \right] 
	+ \frac{g_T}{\sqrt 2} (\partial^2 \tau) (H_u H_d + {\rm h.c.}) .
\end{split}
\end{equation}
Using the equation of motion for the Higgs boson and modulus, we find
\begin{equation}
	\mathcal L = \sqrt{\frac{2}{K_{TT}}} \hat\tau \left[ \left(\frac{\partial_T Z_u}{Z_u}\right) m_{H_u}^2 |\hat H_u|^2+ 
	\left(\frac{\partial_T Z_d}{Z_d}\right) m_{H_d}^2 |\hat H_d|^2 \right] 
	+ \frac{g_T}{\sqrt {2 K_{TT} Z_uZ_d}} m_\tau^2 \hat \tau (\hat H_u \hat H_d + {\rm h.c.}).
\end{equation}
Here the Higgs masses are collectively denoted by $m_{H_u}^2$ and $m_{H_d}^2$.
The second term gives the partial decay width (\ref{tauHH}).
The first terms proportional to $\partial_T Z_u$ and $\partial_T Z_d$ yields the partial width
suppressed by the Higgs masses hence we have neglected it.

\subsection{Moduli-higgsino}

From the higgsino kinetic term, we obtain
\begin{equation}
\begin{split}
	\mathcal L_{\rm kin} =\sqrt{2}(\partial_TZ_{u})\tau \left[ -i  {\tilde H}_u^\dagger \bar\sigma^\mu\partial_\mu \tilde H_u
		+ i (\partial_\mu {\tilde H}_u^\dagger) \bar\sigma^\mu \tilde H_u
	  \right],
\end{split}
\end{equation}
and similarly for $\tilde H_d$.
By using the equation of motion of higgsinos, this becomes
\begin{equation}
	\mathcal L_{\rm kin} =\sqrt{\frac{2}{K_{TT}}} \left(\frac{\partial_T Z_{u}}{Z_u}\right) \mu \hat\tau \hat{\tilde H}_u \hat{\tilde H}_d + {\rm h.c.}
	\label{higgsino_kin}
\end{equation}
where $\mu$ denotes the physical higgsino mass. Here we have assumed that the higgsino-gaugino mixing is negligibly small.

On the other hand, $\mu$ is expressed as~\cite{Wess:1992cp}\footnote{
	If there is an interaction of the Higgs fields with a SUSY breaking field $z$ such as
	$K \supset z^\dag H_u H_d + {\rm h.c.}$, it would contribute to the $\mu$ parameter.
	However, this does not affect the modulus decay into higgsinos, as long as the mixing
	between $T$ and $z$ is negligible.
}
\begin{equation}
	\mu = \frac{1}{\sqrt{Z_uZ_d}}e^{G/2} \nabla_u G_d  =  \frac{1}{\sqrt{Z_uZ_d}}(gm_{3/2} + g_T F^{T*}).
\end{equation}
Then the modulus-higgsino interaction is obtained as
\begin{equation}
\begin{split}
	\mathcal L_{\rm mass} = -\frac{1}{\sqrt{2 K_{TT}Z_uZ_d}}\left[ (2g_T+gK_T)m_{3/2}+ g_T(F^{T*}_T+F^{T*}_{\bar T}) + 2g_{TT}F^{T*}
	\right] \hat\tau \hat{\tilde H}_u \hat{\tilde H}_d +{\rm h.c.}.
	\label{higgsino_mass}
\end{split}
\end{equation}
From (\ref{higgsino_kin}) and (\ref{higgsino_mass}), we obtain the partial decay width (\ref{tauhiggsino}).

\subsection{Moduli-gauge boson}

From the gauge-kinetic function $f_{\rm vis}$, we obtain
\begin{equation}
\begin{split}
	\mathcal L &= -\frac{1}{4\sqrt{2}}\left( {\rm Re} (\partial_T f_{\rm vis}) \tau F_{\mu\nu}^a F^{\mu\nu a} -{\rm Im} (\partial_T f_{\rm vis}) \tau F_{\mu\nu}^a \tilde F^{\mu\nu a}  \right)\\
	&= -\frac{1}{4\sqrt{2K_{TT}}({\rm Re} f_{\rm vis})}\left( {\rm Re} (\partial_T f_{\rm vis}) \hat\tau \hat{F}_{\mu\nu}^a \hat{F}^{\mu\nu a} -{\rm Im} (\partial_T f_{\rm vis}) \hat \tau \hat{F}_{\mu\nu}^a \hat{\tilde F}^{\mu\nu a}  \right)
\end{split}
\end{equation}
From this, we obtain the partial width (\ref{tauAA}).

\subsection{Moduli-gaugino}

From the gaugino kinetic term, we have
\begin{equation}
	\mathcal L_{\rm kin} = -\frac{\tau}{2\sqrt 2}\left[ (\partial_T f_{\rm vis}) (i\lambda^a \sigma^\mu \partial_\mu \bar \lambda^a) +{\rm h.c.} \right],
\end{equation}
which, by using the equation of motion, is rewritten as
\begin{equation}
	\mathcal L_{\rm kin} = \frac{1}{2\sqrt {2K_{TT}}} \frac{\partial_T f_{\rm vis}}{{\rm Re} f_{\rm vis}} m_\lambda \hat\tau  \hat\lambda^a\hat \lambda^a +{\rm h.c.}
	\label{gaugino_kin}
\end{equation}
Here $m_\lambda = (\partial_T f_{\rm vis}) F^T/(2{\rm Re} f_{\rm vis} ) $ is the physical gaugino mass.
From the gaugino mass term, we also have
\begin{equation}
	\mathcal L_{\rm mass} =  - \left[\frac{1}{4\sqrt {2K_{TT}}({\rm Re} f_{\rm vis})}\left\{ (\partial_T f_{\rm vis})(F^T_T+F^T_{\bar T}) + (\partial^2_T f_{\rm vis}) F^T \right\} \hat\tau \hat\lambda^a\hat\lambda^a +{\rm h.c.} \right].
	\label{gaugino_mass}
\end{equation}
From (\ref{gaugino_kin}) and (\ref{gaugino_mass}), we obtain the partial decay width (\ref{taugg}).


\end{document}